\documentclass[12pt]{article}
\usepackage[left=3cm,top=3cm,right=3cm,bottom=2cm]{geometry}
\usepackage{dcolumn}     
\usepackage{multirow}    
\usepackage{rotating}    
\usepackage{graphicx}    
\usepackage{bm}          
\usepackage{amssymb,amsmath}
\usepackage{eso-pic}
\makeatletter
\AddToShipoutPicture*{%
            \setlength{\@tempdimb}{.12\paperwidth}%
            \setlength{\@tempdimc}{.65\paperheight}%
            \setlength{\unitlength}{1pt}%
            \put(\strip@pt\@tempdimb,\strip@pt\@tempdimc){%
        \includegraphics[angle=0,width=6.5in]{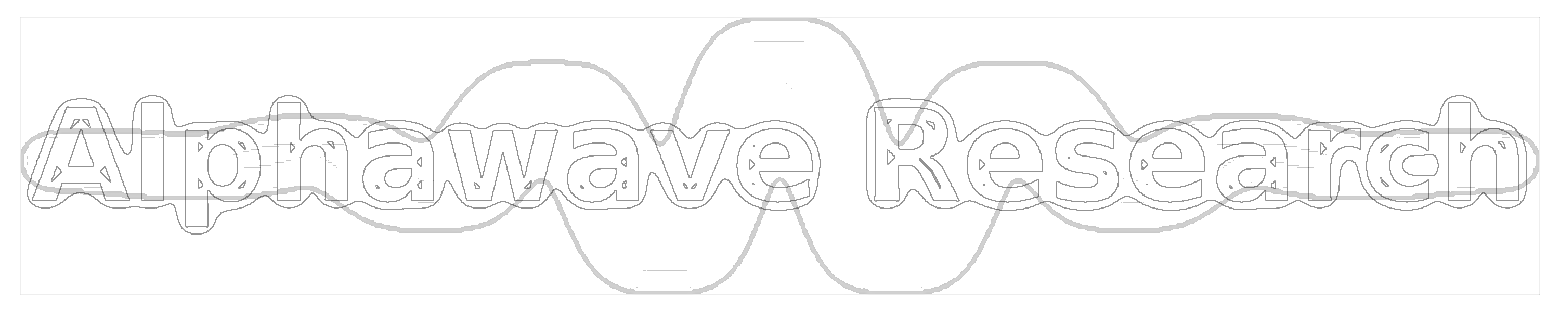}
            }%
}
\makeatother

\author{Robert W. Johnson \\
\small Alphawave Research\\[-0.8ex]
\small Atlanta, GA, USA\\
\small \texttt{robjohnson@alphawaveresearch.com}\\}
\title{Modelling the electric field applied to a tokamak}
\date{\today\\
\small keywords: solenoid, Ohmic heating coil, vector potential, tokamak}
\renewcommand{\vec}[1]{\boldsymbol{#1}}

\newcommand{\beq}{\begin{equation}}
\newcommand{\eeq}{\end{equation}}
\newcommand{\bea}{\begin{eqnarray}}
\newcommand{\eea}{\end{eqnarray}}

\newcommand{\Xhat}{\widehat{\vec{X}}}
\newcommand{\Yhat}{\widehat{\vec{Y}}}

\newcommand{\phihat}{\widehat{\vec{\phi}}}
\newcommand{\Zhat}{\widehat{\vec{Z}}}

\newcommand{\del}{\nabla}
\newcommand{\lap}{\nabla^2}
\newcommand{\divr}{\nabla \cdot}
\newcommand{\curl}{\nabla \times}
\newcommand{\dsub}[1]{\partial_{#1}}

\begin{document}
\maketitle

\begin{abstract}
The vector potential for the Ohmic heating coil system of a tokamak is obtained in semi-analytical form.  Comparison is made to the potential of a simple, finite solenoid.  In the quasi-static limit, the time rate of change of the potential determines the induced electromotive force through the Maxwell-Lodge effect.  Discussion of the gauge constraint is included.
\end{abstract} 
\maketitle
%

\section{Introduction}
The physics of the solenoid~\cite{brick-905B} has recently enjoyed a renewed interest among investigators~\cite{afanas-5755,mcdonald-1176,deolive-1175,batelaan-38} for what it can say about the relationship between, or even the relative reality of, the vector potential and the electromagnetic fields.  While the Aharonov-Bohm effect~\cite{ehrenberg-628,abeffect-1959,abeffect-1961} has received most of the attention, its classical analogue, the Maxwell-Lodge effect~\cite{lodge-469,rouss-49249} suggests that electromotive induction is best described in terms of the potential.

With an aim for developing a model for the electric field applied to toroidal plasma confinement devices, we restrict the scope of this paper to the quasi-static case appropriate for low-frequency phenomena.  The interested reader is directed to the paper by McDonald~\cite{mcdonald-1176} and references therein for a treatment in terms of retarded potentials.

We begin by deriving the vector potential for a simple solenoid of finite length, accounting for every current which one can see.  The model for the circulating potential is built from that obtained for a single current loop.  Then the Ohmic heating coil system of a tokamak is modelled by a circulating potential including contributions from the outer current loops.  The use of field-cancelling outer loops can produce a significant region with no magnetic field present which still experiences electromotive induction.  We conclude with a discussion of the relative merits of the potential and field formulations.

\section{Simple Solenoid}
\begin{figure}
  \includegraphics[scale=.75]{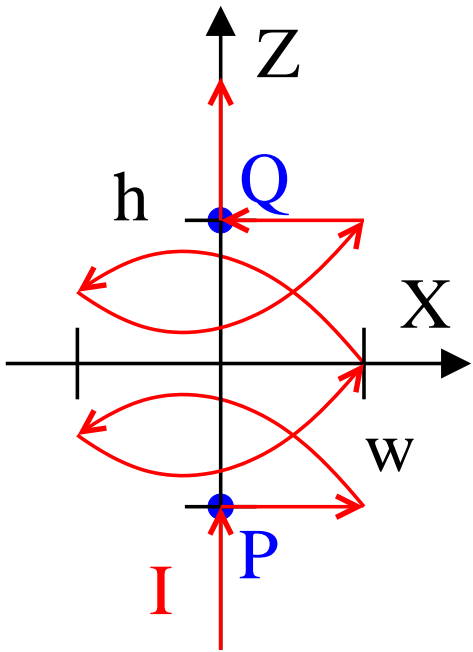}
\caption{Current path for the simple solenoid fed from infinity with horizontal leads at $Y=0$ along $\Xhat$.}
\label{fig:Z}       
\end{figure}

The full expression for the vector potential of a simple solenoid turns out to be not so simple after all!  Once one accounts for the actual path of the current, Figure~\ref{fig:Z}, even with some continuum approximations, the result contains terms which break the axial symmetry (for a single feed wire).  We work in $(Z,X,Y)$ Cartesian and $(Z,R,\phi)$ cylindrical coordinates, where $\dsub{X} \equiv \partial / \partial X$, with the axis of the solenoid along $\Zhat$.  The current $I_0$ is fed from a source at infinity along the $Z$-axis, and the solenoid's spatial extent has a height of $2h$ and a width of $2w$ with $N_h \equiv N/h$ turns per unit length.  The coils are fed from the $Z$-axis by leads at $Y=0$ along $\Xhat$ such that $I_X = I_0$.  Along the coils, the current vector has two components, $\vec{I}_{\mathrm{coil}} = \Zhat I_Z + \phihat I_\phi$ such that $I_\phi / I_Z = 2 \pi w N_h$ and $I_0^2 / I_Z^{\,2} = 1 + (I_\phi/I_Z)^2$.  For $N_h \gg 1$, we make the continuum approximation $\vec{I}_{\mathrm{coil}} \rightarrow \vec{K}_{\mathrm{coil}} = \vec{I}_{\mathrm{coil}} / 2 \pi w$, representing the line current of the conductor by a surface current along the cylinder with both polar and axial contributions.

\begin{figure}
  \includegraphics[scale=.75]{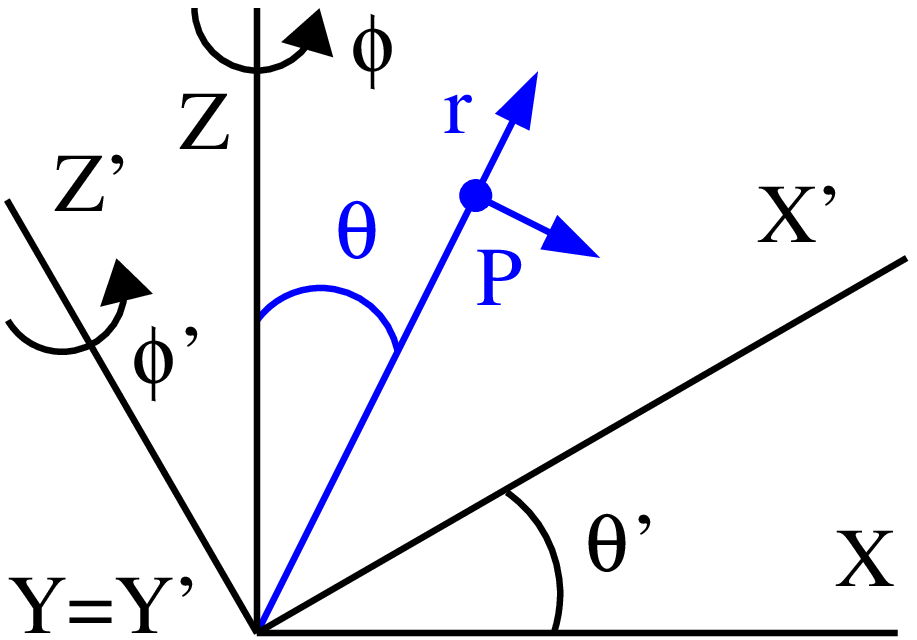}
\caption{Coordinates for a tilted loop normal to $\Zhat'$.}
\label{fig:Y}       
\end{figure}

We note what appears to be the slightest of errors in the geometry used by Rousseaux, {\it et al}~\cite{rouss-49249}, in that the current path along the coil does not constitute closed loops at a fixed inclination to the vertical, which would break the symmetry by the polar angle of the inclination and does not allow for the net flow of charge along the solenoid.  The vector potential of a tilted loop should pick up an additional rotation compared to their expression, as follows.  Let $ \tan \theta' = 1 / N_h 2 \pi R'$ define the pitch angle, as in Figure~\ref{fig:Y}.  A circular loop with radius $R'$ at angle $\theta'$ to the horizontal midplane has $\vert X \vert \leq R' \cos \theta'$, $\vert Y \vert \leq R'$, and $\vert Z \vert \leq R' \sin \theta'$.  The unit vector normal to its area has components $(-\sin \theta',0,\cos \theta')$ in $(X,Y,Z)$ thus breaks the polar symmetry $\partial_\phi \neq 0$ in $(R,\phi,Z)$.  Using $C_\alpha \equiv \cos \alpha$ and similarly for $\sin$ and $(X',Y',Z')$ for vector components in Cartesian coordinates relative to the tilted loop, the spherical components of the potential at a point $P$ are 
\bea
\left[ \begin{array}{c} A_\theta \\ A_\phi \\ A_r \end{array} \right] &=& 
\left[ \begin{array}{ccc} C_\theta & 0 & -S_\theta \\ 0 & 1 & 0 \\ S_\theta & 0 & C_\theta \end{array} \right]
\left[ \begin{array}{ccc} C_\phi & S_\phi & 0 \\ -S_\phi & C_\phi & 0 \\ 0 & 0 & 1 \end{array} \right] \times \\
& & \left[ \begin{array}{ccc} C_{\theta'} & 0 & -S_{\theta'} \\ 0 & 1 & 0 \\ S_{\theta'} & 0 & C_{\theta'} \end{array} \right]
\left[ \begin{array}{ccc} C_{\phi'} & -S_{\phi'} & 0 \\ S_{\phi'} & C_{\phi'} & 0 \\ 0 & 0 & 1 \end{array} \right]
\left[ \begin{array}{c} A_{R'}  \\ A_{\phi'} \\ A_{Z'}  \end{array} \right] \;,
\eea 
relative to a vector on the right giving the potential $\vec{A}_{\mathrm{loop}}$.  On the positive $X$-axis, $\phi=\phi'=0$ and $\vec{A} = \phihat' A_{\phi'} = \Yhat' A_{\phi'} = \phihat A_{\phi'}$, but along the positive $Y$-axis $\phi=\phi'=\pi/2$ and $\vec{A} = \phihat' A_{\phi'} = -\Xhat' A_{\phi'} = -\Xhat A_{\phi'} C_{\theta'} -\Zhat A_{\phi'} S_{\theta'} = \phihat A_{\phi'} C_{\theta'} -\Zhat A_{\phi'} S_{\theta'} = \hat{\theta} A_{\phi'} S_{\theta'} + \phihat A_{\phi'} C_{\theta'}$.

\begin{figure}
  \includegraphics{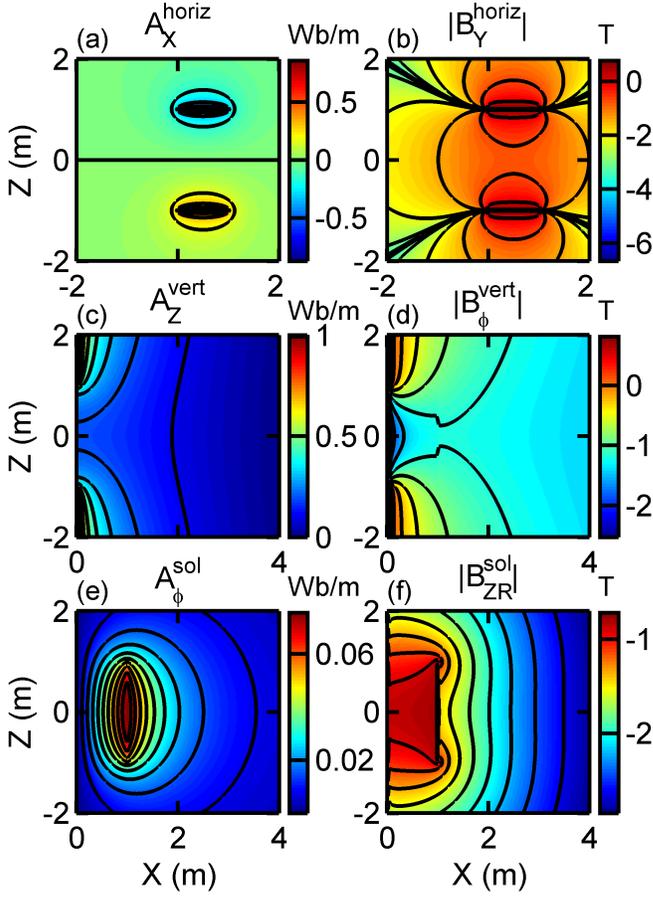}
\caption{Vector potential (left column) and field (right column) produced by a simple solenoid with $h=w=1$m and $N_h = 5$ turns per unit length carrying 1MA of current.  The horizontal contributions (a) and (b) are evaluated at $Y=0$ while the vertical and solenoidal contributions have polar symmetry.  Note that the contours are linearly spaced for $\vec{A}$ and logarithmically spaced (base 10) for $\vec{B}$.}
\label{fig:B}       
\end{figure}

The expression for the magnetic field of an infinitely long, infinitely thin wire is well known, $\vec{B}_{\mathrm{wire}} = \phihat \mu_0 I_0 / 2 \pi R$.  We agree with the position that the symmetry of the potential should match the symmetry of the constituent source, and for a static current density $\divr \vec{J} = 0$ we use the Coloumb gauge constraint $\divr \vec{A} = 0$.  Accordingly~\cite{griffiths-89}, Ampere's law becomes $\lap \vec{A} = - \mu_0 \vec{J}$, with solution (writing $\mu'_o \equiv \mu_0 / 4 \pi$) \beq \label{eqn-poisson}
\vec{A}_{\vec{X}}(\vec{r}) = \mu'_0 \int\!\!\int\!\!\int {\rm d} \tau' \dfrac{\vec{J}_{\vec{X}}(\vec{r}')}{\vert \vec{r}-\vec{r}' \vert}
\eeq for sources of finite extent vanishing to infinity, where the subscript $\vec{X}$ reminds us that only Cartesian components may be extracted from the integral directly, and similarly for surface and line currents $\vec{K} = \delta_\sigma \vec{J}$ and $\vec{I} = \delta_\lambda \vec{K}$.  Sources which extend to infinity require various ``tricks of the trade'' for the determination of their potential; for the infinite wire, inspection reveals that $\vec{A}_{\mathrm{wire}} = - \Zhat \mu'_0 I_0 \ln R^2$ satisfies $\curl \vec{A}_{\mathrm{wire}} = \vec{B}_{\mathrm{wire}}$.  For our current path there is a missing segment of current between points P and Q in the figure, whose potential is written \bea
\vec{A}_{\mathrm{seg}}^h &=& \Zhat \mu'_0 I_0 \int_{-h}^h {\rm d} Z' [(Z-Z')^2 + R^2]^{-1/2} \;, \\
 &=& \Zhat \mu'_0 I_0 \ln \left \lbrace \dfrac{h-Z+[(Z-h)^2+R^2]^{1/2}}{-h-Z+[(Z+h)^2+R^2]^{1/2}} \right \rbrace \;,
\eea and one may recover the infinite wire's expression\footnote{\textit{cf} J. B. Tatum, \textit{Physics Topics: Electricity and Magnetism} at http://www.astro.uvic.ca/$\sim$tatum/elmag.html } by doubling the semi-infinite integral for the potential in the plane $Z=0$ given by taking the limit $h \rightarrow \infty$ and dropping the infinite constant's contribution, \bea
\vec{A}_{\mathrm{seg}}^\infty &=& \Zhat \mu'_0 I_0 2 \ln \left [ Z' + ({Z'}^{\,2} + R^2)^{1/2} \right ] \biggr \vert_0^\infty \\
&\rightarrow& - \Zhat \mu'_0 I_0 \ln R^2 = \vec{A}_{\mathrm{wire}} \;.
\eea  For the horizontal feeds, one has the axial symmetry breaking terms \bea
\vec{A}_{\mathrm{horiz}} &= &\Xhat \mu'_0 I_0 \times \\ & &\sum_{Z'} \ln \left \lbrace \dfrac{w-X+[(X-w)^2+Y^2+(Z-Z')^2]^{1/2}}{-X+[X^2+Y^2+(Z-Z')^2]^{1/2}} \right \rbrace \nonumber
\eea for $Z' = \pm h$, evaluated for the plane along the $X$ axis in Figure~\ref{fig:B} (a) with its field in (b), where the potential changes sign at the horizontal midplane and the vertical component of the field vanishes at $Y=0$.

Next we tackle the vertical current flowing along the coils at the coil radius $R' = w$, which is equivalent to a surface current flowing along a finite, hollow pipe $\vec{K}_{\mathrm{pipe}}$.  It produces a potential with axial symmetry, \beq
\vec{A}_{\mathrm{pipe}}^{h, R'} = \Zhat \mu'_0 \dfrac{I_Z}{2 \pi R'} \int_{-\pi}^\pi {\rm d} \phi' R' \int_{-h}^h \dfrac{{\rm d} Z' }{\vert \vec{r}-\vec{r}' \vert}
\eeq for $\vert \vec{r}-\vec{r}' \vert^2 = (Z-Z')^2 + R^2 +{R'}^{\,2} - 2 R R' \cos (\phi - \phi')$, which we wish to compute along the $X$-axis where $\phi=0$.  For either order of integration, while the first is tractable, the second resists being put into closed form and requires a numerical evaluation.  For the infinite pipe $h \rightarrow \infty$, its integral \bea
\vec{A}_{\mathrm{pipe}}^{\infty, R'} &=& - \Zhat \mu'_0 \dfrac{I_Z}{2 \pi} \int_{-\pi}^\pi {\rm d} \phi' \ln \left ( R^2 +{R'}^{\,2} - 2 R R' \cos \phi' \right ) \;, \\
&=& - \Zhat \mu'_0 \dfrac{I_Z}{2 \pi} \int_{-\pi}^\pi {\rm d} \phi' \ln (a - 2 b \cos \phi')
\eea may be parametrized by $a=R^2 +{R'}^{\,2}$ and $b=R R'$, where $a \geq 2 b$ with equality at the conductor.
A direct evaluation thereof requires attention to the logarithmic branch which previously has led us into difficulties.  Avoiding some complex analysis, we expand the integrand using $c = 2 b / a$, \bea
\ln (a - 2 b \cos \phi') &= \ln a + \ln (1 - c \cos \phi') \;, \\
&= \ln a - \sum_{k=1}^\infty (c \cos \phi')^k / k \;,
\eea and integrate under the sum for non-vanishing even powers, \bea
\int_{-\pi}^\pi {\rm d} \phi' \ln (1 - c \cos \phi') &= - \sum_{k=1}^\infty \dfrac{}{}\!\!\! \int_{-\pi}^\pi {\rm d} \phi' \dfrac{(c \cos \phi')^k}{k} \;, \\
&= - \sum_{k=1}^\infty \dfrac{c^k}{k} \int_{-\pi}^\pi {\rm d} \phi' (\cos \phi')^k \;, \\
&= - \sum_{k=1}^\infty \dfrac{c^{2k}}{2k} 2 \beta \left( \dfrac{1}{2}, \dfrac{2k+1}{2} \right) \;,
\eea where the beta function $\beta (x, y) = \Gamma(x) \Gamma(y) / \Gamma(x+y)$ may be expressed~\cite{Press:1992,abramowitz-stegun} in terms of the gamma function, to obtain the result \beq
\vec{A}_{\mathrm{pipe}}^{\infty, R'} = - \Zhat \mu'_0 I_Z \left[ \ln a - \dfrac{1}{2\pi} \sum_{k=1}^\infty \dfrac{c^{2k}}{k} \beta \left( \dfrac{1}{2}, k+\dfrac{1}{2} \right) \right] \;.
\eeq  

\begin{figure}
  \includegraphics{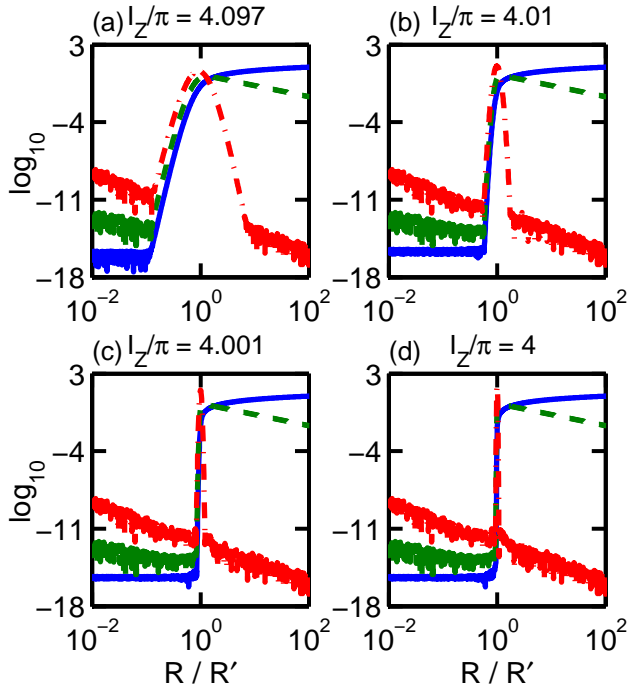}
\caption{Vector potential (solid), magnetic field (dashed), and current density (dash-dot) for an infinite pipe at increasing order of evaluation: (a) $k_{max}=10$, (b) $k_{max}=100$, (c) $k_{max}=1000$, and (d) $k_{max}=10,000$.  The equivalent current on axis equals $4\pi$.}
\label{fig:A}       
\end{figure}

With normalization $\mu_0 \equiv 1$ and $I_Z = 4 \pi$, inspection of the potential as the count of terms $k_{max}$ increases, Figure~\ref{fig:A}, reveals that one may write \beq
\vec{A}_{\mathrm{pipe}}^{\infty, R'} = - \Zhat H (R/R'-1) 2 \log (R/R') \;,
\eeq where $H(x)$ is the Heaviside step function.  Its curl $\vec{B}_{\mathrm{pipe}}^{\infty, R'} = - \phihat \dsub{R} A_Z$ gives the field \bea
\vec{B}_{\mathrm{pipe}}^{\infty, R'} &= &\phihat \dfrac{2}{R R'} \left[ R' H(R/R'-1) \right. \\
& &\left. + R \delta(R/R'-1) \log (R/R') \right] \;,
\eea where the second term vanishes on account of the Dirac function $\delta(x) = \vert a \vert \delta(ax)$, and its Laplacian is \beq
\del^2 A_Z = - \dfrac{2}{R R'} \delta(R/R'-1) = - \dfrac{2}{R} \delta(R-R')\;,
\eeq whence the surface current equals $K_Z = 2 / R'$ recovering $2 \pi R' K_Z = 4 \pi = I_Z$.  For the finite pipe, one may verify that \beq \label{eqn:hpipe}
\vec{A}_{\mathrm{pipe}}^{h, R'} = \sum_{-\pi}^\pi \Delta \phi' R' \vec{A}_{\mathrm{seg}}^{R', \phi'} = \sum_{-h}^h \Delta Z' \vec{A}_{\mathrm{ring}}^{Z', R'}
\eeq give equivalent evaluations in the limit $\Delta \phi', \Delta Z' \rightarrow 0$, where \bea
\vec{A}_{\mathrm{seg}}^{R', \phi'} &= &\Zhat \mu'_0 \dfrac{I_Z}{2 \pi R'} \times \\ & &\ln \left \lbrace \dfrac{h-Z+[(Z-h)^2 + a - 2 b \cos \phi']^{1/2}}{-h-Z+[(Z+h)^2 + a - 2 b \cos \phi']^{1/2}} \right \rbrace \;, \nonumber
\eea and $\vec{A}_{\mathrm{ring}}^{Z', R'}$ will be given later.  When adding up these contributions for the solenoid, $\vec{A}_{\mathrm{vert}} = \vec{A}_{\mathrm{wire}} - \vec{A}_{\mathrm{seg}} + \vec{A}_{\mathrm{pipe}}$ in  Figure~\ref{fig:B} (c), one finds the relative magnitude $A_\mathrm{pipe} / A_\mathrm{seg}$ is reduced by both the pitch angle $N_h$ and the circumference $2 \pi R'$.  The corresponding field in (d) just barely shows the influence of $\vec{A}_{\mathrm{pipe}}$ beyond the radius $w$ even for a contrivedly low turns per length of $N_h = 5$.

Finally comes the potential arising from the solenoidal current, evaluated at $\phi=0$ with contributions from the $Y$ component of the coil current $\vec{K}_\phi$, \beq
\vec{A}_{\mathrm{sol}} = \phihat \mu'_0 \dfrac{I_\phi}{2\pi} \int_{-h}^h {\rm d} Z' \int_{-\pi}^\pi \dfrac{{\rm d} \phi' \cos \phi'}{\vert \vec{r}-\vec{r}' \vert}
\eeq for $\vert \vec{r}-\vec{r}' \vert^2 = a-2 b \cos \phi'$, where now $a=(Z-Z')^2+R^2+{R'}^{\,2} \geq 2b$ with equality at the location of the loop.  The angular integral evaluated for a single current loop at $Z'$ may be expressed in terms of complete elliptic integrals with parameter $k^2 = - 4 b / (a-2b)$ as \bea
A_{\mathrm{loop}}^{Z', R'} &\propto& \int_{-\pi}^\pi \dfrac{{\rm d} \phi' \cos \phi'}{\sqrt{a-2 b \cos \phi'}} \;, \\
 &=& \dfrac{2}{b\sqrt{a-2b}} \left [ a K(k) - (a - 2 b) E(k) \right ] \;,
\eea whereupon writing $k \rightarrow {\rm i} k$ and $k' = \sqrt{1-({\rm i} k)^2}$ produces a real modulus $k'' = k/k'$, and the elliptic integrals transform~\cite{abramowitz-stegun} as $K({\rm i} k) = K(k'') /k'$ and $E({\rm i} k) = E(k'') k'$, yielding \bea
A_{\mathrm{loop}}^{Z', R'} &\propto& \dfrac{2}{b\sqrt{a+2b}} \left [ a K(k'') - (a + 2 b) E(k'') \right ] \;, \\
 &=& \dfrac{2 \sqrt{a+2b}}{b} \left [ \left( \dfrac{a}{a + 2 b} \right) K(k'') -  E(k'') \right ] \;,
\eea for $k''=\sqrt{4b/(a+2b)}$,  which compares well with other derivations~\cite{deolive-1175,Laslett-1987bq} following examples by Jackson \cite{jackson-third} and Smythe~\cite{Smythe-887954}, noting that the current around the infinitesimal loop differs from the current around one of the solenoidal loops by a factor $I'_\phi = \delta_{Z'} K_\phi = \delta_{Z'} I_\phi / 2 \pi R'$.  The remaining integral over $Z'$ has so far eluded capture in closed form; however, its numerical evaluation on a rectangular grid is simply accomplished by adding up shifted copies of the loop potential with the appropriate normalization, $\vec{A}_{\mathrm{sol}} = \sum_{Z'} \Delta Z' \vec{A}_{\mathrm{loop}}^{Z', R'}$, for inter-loop spacing $\Delta_{Z'}$.  The summation results in the potential and field shown in Figure~\ref{fig:B} (e) and (f), where the sign of $B_Z$ changes from the inner to outer region and $B_R$ vanishes on the horizontal midplane.  By a similar calculation, one obtains the potential for a ring of current appearing in Equation~(\ref{eqn:hpipe}), \beq
\vec{A}_{\mathrm{ring}}^{Z', R'} = \Zhat \mu'_0 \dfrac{I_Z}{2\pi} \dfrac{4}{\sqrt{a+2b}} K(k'') \;.
\eeq  The final expression for the vector potential of a simple solenoid is then written as $\vec{A}_{\mathrm{simple}} = \vec{A}_{\mathrm{horiz}} + \vec{A}_{\mathrm{vert}} + \vec{A}_{\mathrm{sol}}$.

\begin{figure}
  \includegraphics[scale=.5]{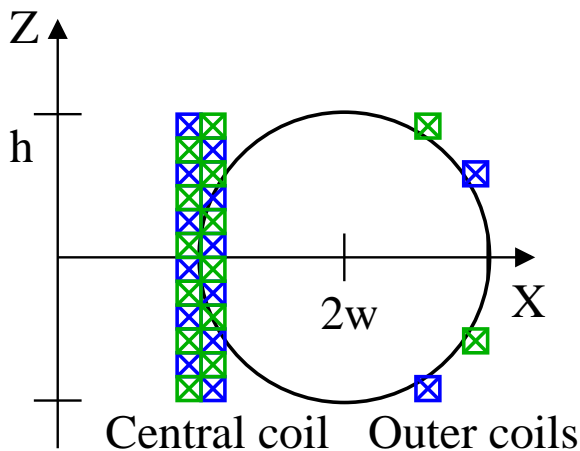}
\caption{Coil locations of a typical Ohmic heating coil assembly.  Two sets of coils in a mirrored configuration are connected in parallel to the same current source.  The outer coils serve to cancel the field within the confinement region.}
\label{fig:X}       
\end{figure}

\section{Ohmic Heating Coil}
The Ohmic heating coil system of a typical tokamak~\cite{diiid-2002} consists of a central solenoid connected in series ({\it ie} with the same current) to outer loops designed to exclude the magnetic field from the confinement region, shown schematically in Figure~\ref{fig:X}.  The central solenoid often consists of a doubled coil fed by a bi-directional cable, thus the vertical and horizontal feed currents effectively cancel, leaving only the solenoidal contribution.  Details of the connections to and between the outer loops have not yet been provided, so we must here neglect their effect.  We model the outer loops by two pairs of coils with vertical symmetry located at $R'=2.5w$ and $3w$ and $Z'=\pm h$ and $\pm h/2$ with $N_1=8$ and $N_2=11$ turns, respectively, and neglect their spatial extent.  The number of turns has been selected to minimize the integral of the magnetic field's magnitude over the confinement region, which we approximate with a circular vessel of radius $w$ centered at $R=2w$.  The magnitude of the current swing can be estimated as $\pm 110$kA per second~\cite{diiid-2002}, with a pulse duration exceeding several seconds.  The electric field distribution on the order of mV/m may be read directly from the vector potential in the quasi-static approximation $\vec{I} \rightarrow \dsub{t}\vec{I}$ as $\vec{E} = - \dsub{t} \vec{A}$ when $\del \Phi=0$.

\begin{figure}
  \includegraphics{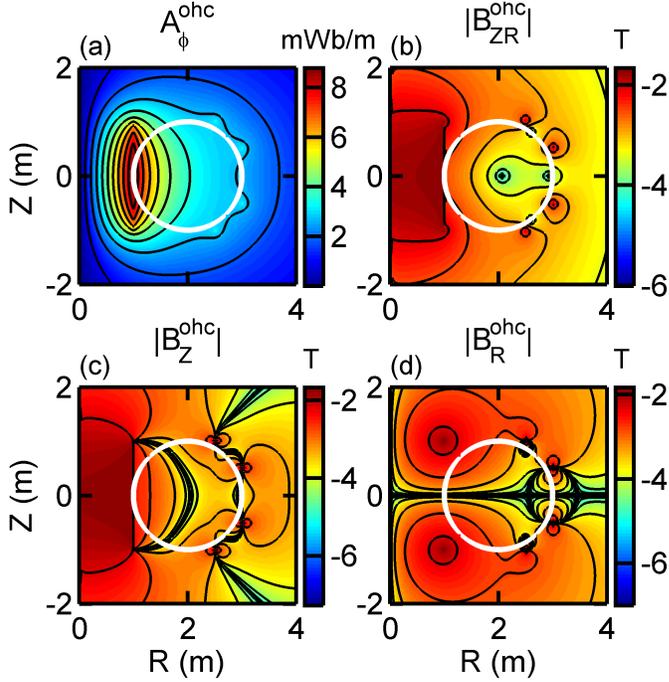}
\caption{Potential (a) and field (b)-(d) produced by the solenoidal component of a typical Ohmic heating coil system carrying 110kA of current with $h=w=1$m, $N_h=50$, and two pair of outer coils.  The white circle represents the confinement vessel, at whose center the field vanishes but the potential does not.  Note that the contours are linearly spaced for $\vec{A}$ and logarithmically spaced (base 10) for $\vec{B}$.}
\label{fig:C}       
\end{figure}

The presence of the outer coils serves to flatten the potential as shown in Figure~\ref{fig:C} (a), thus reducing the magnetic field within the confinement region, as seen in (b).  Both the vertical (c) and horizontal (d) field components change sign within the vacuum vessel, and for a sufficiently complicated outer coil configuration, a field-free region of significant extent can be achieved.  The exclusion of the field in no way affects the electromotive induction present at the center of the vessel resulting from the time rate of change of the vector potential $\dsub{t}\vec{A}_{\mathrm{ohc}}$, which does {\it not} vanish within the confinement region.  The situation begs the question, how does a charged particle know that it should move when at the location of vanishing magnetic field?  We feel the answer lies in ascribing to the classical potential a sense of reality which surpasses that of the field formulation.

\begin{figure}
  \includegraphics{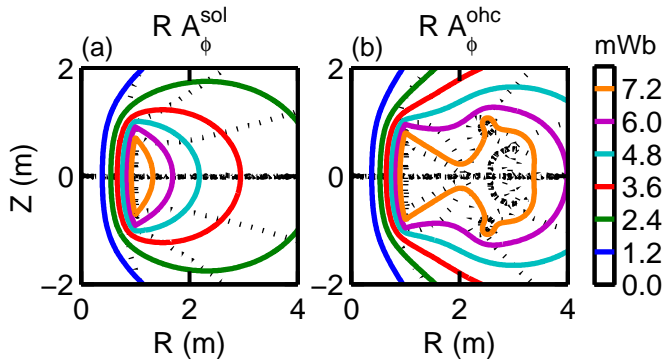}
\caption{Contours (solid) of the flux function $\maltese \psi = R \vec{B}_{ZR}$ for $-\psi = R \vec{A}_\phi$ give the field lines for the central solenoid (a) and the Ohmic heating coil system (b).  Also shown are contours (dotted) of the field line's angle evenly spaced in units of $\pi/4$.}
\label{fig:D}       
\end{figure}

We close this section by noting that the contours of $A_\phi$ do not give directly what would be called the flux surfaces for this configuration.  For $R \divr \vec{B}_{ZR} = 0$ with $\dsub{\phi} = 0$, the field $\vec{B}_{ZR}$ may be written as the contours of a flux function $(-\dsub{R}, \dsub{Z}) \psi = \phihat \times \del \psi \equiv \maltese \psi$ such that $\maltese \psi = R \vec{B}_{ZR} = R \curl \vec{A}_\phi$, thus $\del (R A_\phi + \psi) = 0$ and the flux function is given (up to an unphysical constant) by $-R A_\phi$.  The central solenoid's contribution is shown in Figure~\ref{fig:D} (a) with the field's angle from the vertical computed directly from $\tan \theta = B_Z / B_R$.  Compared to the flux function of the simple solenoid, the Ohmic heating coil system (b) has a much more complicated topography induced by the outer coil assembly.

\section{Discussion and Conclusion}
The argument over which is more fundamental, the potential or the field, has yet to be decided~\cite{maxwell-1865,lodge-book,feynmanlecs,fried-2835F,roche-22291,jackson-917,rousseaux-2005-30}.  From the hydrodynamical analogue~\cite{rouss-00013234}, one sees that the choice of (Coulomb)Lorenz gauge reflects a statement on the (in)compressibility of the electromagnetic potential/flow and on whether the speed of propagation of disturbances is (in)finite.  Accordingly, in situations when $\divr \vec{J}=0$, one should take the Coulomb gauge $\divr \vec{A}=0$, and when a varying space charge density may develop $\dsub{t} \rho_e \neq 0$, then the Lorenz gauge $\dsub{\mu} A^\mu=0$ is appropriate.  From the geometric statement~\cite{ryder-qft,davis70,naka-798212,ward-1286287} of classical electromagnetism in terms of dual forms ${\rm d}^*{\rm d} A = J$, an intimate relationship is apparent between the potential and the source.  We feel that this relationship is beyond cosmetic and speaks to the fundamental structure of natural phenomena.

In executing the solution of the Poisson equations~(\ref{eqn-poisson}), one evaluates at a point $(Z,X,Y)$ the weighted contribution from all source points $(Z',X',Y')$ which have had time to communicate their influence to the point in question, which for the Coulomb gauge with infinite propagation speed of disturbances extends to infinity.  Conversely, the action of the current source has made an image of itself which disperses through the region of consideration at the speed of disturbances, which we call the potential.  The description in terms of diverging and circulating fields adds a level of abstraction analogous to ascribing to the slope of a hill the same reality as the hill itself; while the slope of the hill informs a ball how to roll under gravity, it is the material of the hill which is real.  Note that we are not advocating a revival of the luminiferous aether~\cite{PRL-020401}, but simply a more fundamental view of the nature of the potential.  Its supposed indeterminacy reflects the fact that a particular set of fields within a region can be generated by a multitude of source configurations external to or within that region; however, once the source has been specified, then so has the shape of the potential, hence the fields.  Just because we cannot see it does not mean it is not there.

Operation of a tokamak (or similar, smaller apparatus~\cite{rouss-49249}) affords the opportunity to explore phenomena beyond that associated with fusion and to think about issues of basic science, such as the nature of induction.  The description of complicated current configurations becomes straightforward in the potential formulation.  Examination of the vector potential applied to a tokamak indicates that it is a prime example of the Maxwell-Lodge effect, where one tries to understand how a charge knows it is to move when there is no field changing with time at its location.

Attempts to ascertain what is happening within the device~\cite{frc-pop-2006} would do well to begin by determining what is being done to the device.  The true starting point of any calculation involving tokamak operation is the determination of the potential produced by sources external to the confinement region.  The usual evaluation in terms of resistive magnetohydrodynamics utilizing the quasi-neutral approximation requires the electric field be determined from an equation of motion rather than Poisson's equation for the potential, thereby denying any predictive use of the model.  From a determination of the applied potential, one should in principle use the Ohm's law equation (conspicuous by its absence beyond the kinetic term in our treatment~\cite{rwj-jpp03} of the macroscopic field formulation) to predict the resulting current density.  Only then can one treat both the fluid and the potential through a hydrodynamic field theory.  The goal of experimental science is to predict results, not to interpret them, and for a tokamak that requires modelling the electric field which the investigator has applied.


\section*{Acknowledgments}
The author appreciates occasional conversations with Ian Aitchison and Germain Rousseaux on the nature of field theory and with various colleagues who caught an early inconsistency.

%

\end{document}